# Understanding collective human movement dynamics during large-scale events using big geosocial data analytics

Junchuan Fan[a,*], Kathleen Stewart[b]

[a] *Oak Ridge National Laboratory, United States*
[b] *Center for Geospatial Information Science, University of Maryland, College Park, United States*

## 1. Introduction

Conventional approaches for modeling human movement dynamics often focus on activity and mobility patterns in individuals' regular daily lives (González, Hidalgo, and Barabási 2008; C. Song et al. 2010). When there are large-scale external events, traditional models may fall short in capturing changes in human movement dynamics in response to large-scale events such as earthquakes (Lu, Bengtsson, and Holme 2012; Bagrow, Wang, and Barabási 2011; X. Song et al. 2017), hurricanes (Wang and Taylor 2014; Roy, Cebrian, and Hasan 2019), epidemic outbreaks (Balcan et al. 2009; Eubank et al. 2004), and major sporting event (Marques-Neto et al. 2018). While it is critically important to understand the impact of large-scale external events on collective movement dynamics, especially for application domains such as forecasting transportation demand and emergency response, the lack of large-scale mobility data during a large-scale event poses significant challenges for researchers. With the rapid advancement of information and communication technologies, many researchers have adopted alternative data sources (e.g., cell phone records, GPS trajectory data) from private data vendors to study human movement dynamics in response to large-scale natural or societal events.

With the ubiquity of location-aware mobile devices, big geosocial data, i.e., socially sensed data with georeferenced information, actively contributed by massive numbers of human sensors have also become a valuable data source for large-scale studies of human behavior (Ruths and Pfeffer 2014). Compared with other mobile device data sources such as cell phone records and GPS logs, *geosocial data* such as georeferenced tweets have their unique characteristics. On the one hand, they are publicly available and dynamically evolving as real-world events are happening, and users' proactive contributions make geosocial data more likely to capture the real-time sentiments and responses of populations. On the other hand, precisely-geolocated geosocial data is scarce and biased toward urban population centers due to people's privacy concerns and population distribution (Johnson et al. 2016). Proper data augmentation and analytic methods have to be developed to mitigate the data scarcity and sampling bias issue of georeferenced tweets in order to mine human activity and movement dynamics.

The 2017 Great American Eclipse was a once-in-decades total solar eclipse that passed from the Pacific to the Atlantic coasts. The total eclipse began in Oregon at 10:16 am PDT and ended in South Carolina at 2:44pm EDT; as the total eclipse moved from the west coast to the east coast, it was visible within a 70 miles wide path across the entire contiguous United States, referred to as the path of totality (POT). With 12 million people living in the path of totality (POT) and about 200 million people living within a day's drive of the POT, hundreds of thousands of people were estimated to have travelled to the POT to watch the total eclipse. On the day of the solar eclipse (Aug 21, 2017), many interstate highways leading to the POT experienced unprecedented traffic congestion, lasting hours. This event also created unprecedented challenges for many locations in the POT. For example, more than 100,000 people flocked to Madras, Oregon, a small town of 7,000 on the eclipse day, and the Oregon National Guard had to be called in to help with logistics and crowd control[1]. Uncovering human movement dynamics in response to such a large-scale event can be extremely helpful for transportation planning, city management, or even emergency response. The wide coverage and real-time response characteristics of Twitter data make it a data source with a unique potential for this task. For this research, we developed a big geosocial data analytical framework for extracting human movement dynamics in response to large-scale events from publicly available georeferenced tweets. The framework includes a two-stage data collection module that collects data in a more targeted fashion in order to mitigate the data scarcity issue of georeferenced tweets; in addition, a variable bandwidth kernel density estimation (VB-KDE) module was developed to fuse georeferenced information at different spatial scales, further augmenting the signals of human movement dynamics contained in georeferenced tweets. The spatial scale of collected georeferenced tweets ranges from *poi to country* (i.e., *country*, *admin* (i.e., state), *city*, *neighborhood*, and *poi*. To correct for the inherent sampling bias of georeferenced tweets, we adjusted the number of tweets for different spatial units (e.g., county, state) by population. Given the social nature of geosocial data, population is an important factor for the adjustment of tweet data sample, or geosocial data generally. There could be other sensible ways to further adjust the tweet data sampling bias depending on the specific event under study. To demonstrate the performance of the proposed analytic framework, we chose an astronomical event that occurred nationwide across the United States, i.e., the 2017 Great American Eclipse, as an example event and studied the human movement dynamics in response to this event. However, this analytic framework can easily be applied to other types of large-scale events such as hurricanes or earthquakes. Specifically, in this research, we address the following research questions: 1) *To what extent can big geosocial data capture human movement dynamics in response to a large-scale event;* 2) *What methods can be used to mitigate the widely acknowledged data sampling bias and data scarcity issues of geosocial data in order to extract human movement dynamics from big geosocial data*; and 3) *How can geospatial data be fused at different spatial scales to augment the signals of human movement dynamics*?

---

[1] https://www.orlandosentinel.com/la-sci-great-american-eclipse-liveblog-it-was-so-busy-in-madras-oregon-they-1503273850-htmlstory.html

The rest of the paper is organized as follows: section 2 discusses related work; section 3 discusses different components of the proposed big geosocial data analytic framework, including the two-stage data collection process and the VB-KDE module; as a validation, section 4 compares the estimation results with traffic count reports released by State Departments of Transportation (DOT); section 5 and 6 discusses the results and presents conclusions and future work.

## 2. Related work

Researchers traditionally uses traffic count data or household travel survey (e.g., the National Household Travel Survey) to track population movement dynamics. Recently, as location-enabled devices become more common, different alternative data sources have been adopted to study human activity patterns and movement dynamics. For example, mobile phone records of millions of users were used to analyze the population movements after the 2010 Haiti earthquake (Lu, Bengtsson, and Holme 2012),  to examine the real-time changes in communication and mobility pattern in the vicinity of eight emergency events (Bagrow, Wang, and Barabási 2011), and to study the spatiotemporal dynamics of movement patterns of participants of large-scale event and the resulting dynamics in the mobile network workload (Marques-Neto et al. 2018). GPS records collected from mobile phones have been used to predict and simulate human mobility following natural disasters (X. Song et al. 2017). GPS trajectory data coupled with contextual information have been used to analyze human mobility and traffic patterns (Siła-Nowicka et al. 2016; Fan et al. 2019).

Most of these data sources can only be accessed with the collaboration of private data owners. Socially sensed data, with its public availability and real-time nature, offers researchers a unique opportunity for large-scale studies of human behavior (Ruths and Pfeffer 2014). Researchers have used socially sensed data to study population response to important events, be it unexpected disaster events or planned events (Guan et al. 2014; Brambilla et al. 2017). Geosocial data was used to estimate different aspects of human mobility, such as individual movement orbits and travel modes (Jurdak et al. 2015), the uncertainty in regular human mobility patterns (Huang and Wong 2015), and the impact of extreme disaster events such as hurricane, earthquake on human mobility resilience (Roy, Cebrian, and Hasan 2019). Researchers also studied the feasibility of using large-scale Twitter data as a proxy of human mobility for modeling and predicting disease spread (Liu et al. 2015).

Using Twitter data to study human movement dynamics often requires inferring users' home locations. Various approaches have been designed to determine Twitter users' home locations based on available information, including using self-reported location in a user's profile (Poblete et al. 2011; Nguyen et al. 2017), calculating the geometric median of a user's tweet locations (Jurgens et al. 2015), finding the region in which a user has posted the most tweets (McNeill, Bright, and Hale 2017), or identifying the region from which a user has posted tweets at least $n$ days apart (Kulshrestha et al. 2012; Hecht and Stephens 2014). Since tweets with associated geographic coordinates are not as common due to changes in how mobile devices capture and track movements, previous research has also explored various different ways to infer the geographic context of tweets without relying solely on geographic coordinates provided by users. For example, researchers have tried to infer Twitter users' geographic locations based on the locations of users' friends (Kong, Liu, and Huang 2014; McGee, Caverlee, and Cheng 2013). Based on the idea that users' tweets may encode some location-specific content, Cheng et al. proposed a probabilistic framework for estimating Twitter users' city-level locations based on the content of users' tweets (Cheng, Caverlee, and Lee 2010). The performances of these algorithms, however, did not scale to real-world conditions (Jurgens et al. 2015). Kernel density estimation is a classic approach for estimating home range of moving entities (Van Winkle, 1975; Worton, 1987). Bandwidth selection is an important parameter for kernel density estimation. Adaptive kernel density estimation method can vary the bandwidth according to the density of data points, using a narrower kernel for areas with high point density and broader kernel for areas with sparse data (Salgado-Ugarte & Pérez-Hernández, 2003; Silverman, 1986). This method was applied on geotagged tweets to construct catchment areas of retail centers (Lloyd & Cheshire, 2017). Essentially, the goal of adaptive kernels is to better account for data uncertainty. Given the multi-scale georeferenced tweets from a Twitter user, this research varies the kernel bandwidth according to spatial scale of the tweet instead of density of the tweets.

## 3. Analytic framework for estimating human movement dynamics from big geosocial data

In this research, we developed a big geosocial data analytic framework, *TweetMovementFlow*, to extract human movement dynamics in response to large-scale events from georeferenced tweets that includes four modules (Figure 1). The first module *identifies and collects socially sensed signals in response to the event.* For the solar eclipse event, *eclipse tweets*, i.e., tweets that include either the keywords '*eclipse*' or '*totality*', were collected during the week of the solar eclipse. The second module creates *a sample of social sensors based on collected data.* For the solar eclipse event, a collection of *POT eclipse watchers,* that is, Twitter users who have posted eclipse tweets (referred to as *eclipse watchers*) inside the POT were identified as our sample for estimating the human movement pattern from different states to *POT states,* i.e., states that intersected with the POT. The third component infers *home location of sampled social sensors* in order to estimate the origin and destination of the eclipse-watching trips made by all sampled social sensors. Since the tweets collected in the first module was only a one-week snapshot, they were too scarce to estimate eclipse watchers' home state. The second stage of data collection process, i.e., retrieving tweet histories of POT eclipse watchers, was developed in the third module. POT eclipse watchers' home states were inferred based on their tweet histories.

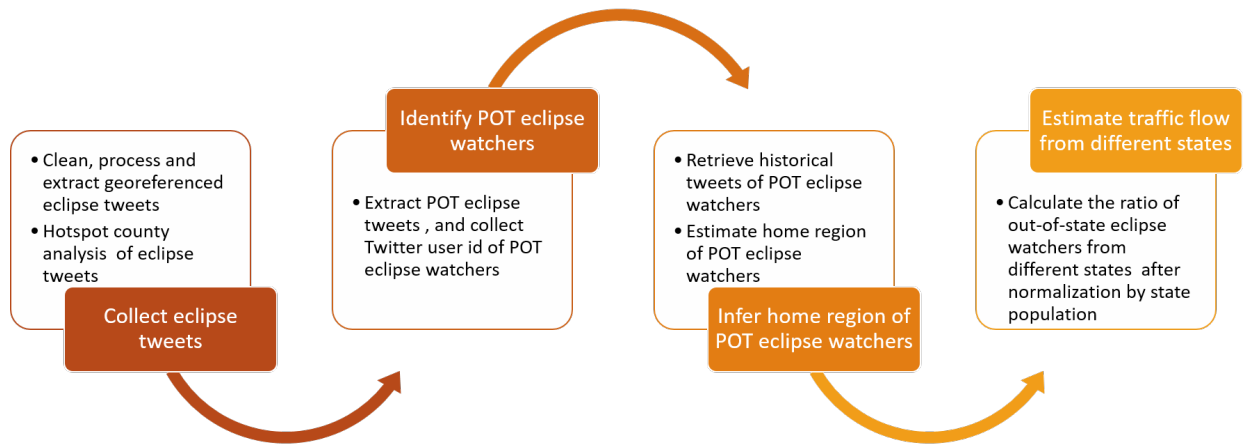

*Figure 1 Analytical framework for estimating human movement pattern to different POT states using georeferenced tweets*

The fourth component estimates *aggregated movement patterns after adjusting the data sampling bias*. The fourth component aggregated the number of eclipse watchers by state and adjusted the number by state population to estimate the percent of visitors from different states. Although this analytic framework is presented in the context of the solar eclipse event, it can easily be adapted to other types of large-scale events.

## 3.1 Spatiotemporal characteristic of eclipse tweets

In the first stage of the data collection process for this study, the *location filter* of the Twitter Streaming API was used to collect only the georeferenced tweets, including both *place-referenced* tweets and *coordinate-referenced* tweets, during the week of the Great American Solar Eclipse (18/08/2017~08/25/2017). A *place-referenced* tweet has a place boundary representing the location of tweet, with five different spatial scales, namely, *country*, *admin* (i.e., state), *city*, *neighborhood*, and *poi*. A *coordinate-referenced* tweet has a pair of geographic coordinates representing the location of tweet. There were 13,908,147 georeferenced tweets, out of which 331,415 were *eclipse tweets*, posted by 241,282 *eclipse watchers*.

The spatial distribution of eclipse tweets collected during the eclipse week (Figure 2a) showed that eclipse tweets were posted from all over the U.S. during the eclipse week, and urban population centers had more coverage. Hotspot analysis was carried out for U.S. counties based on the number of eclipse tweets in each county. Eclipse tweets with coarser than county-scale georeferenced information were excluded from hotspot analysis (Figure 2(b) (c)). After excluding eclipse tweets that were at country or state scale, there were 284,462 (85.8%) eclipse tweets with city or finer-than-city-scale georeferenced information. The number of eclipse tweets in each U.S. county was then counted by spatially joining place boundary center or coordinate of eclipse tweets with U.S. counties. Figure 2b shows the result of hotspot county analysis for eclipse tweets. Without correction for tweet data sampling bias, the hotspot counties were mostly large urban population centers in the U.S. with high population density, reflecting tweet data's spatial sampling bias. After normalizing the number of eclipse tweets in each county by the county population, the hotspot county results clearly revealed the effect of the solar eclipse event (Figure 2c). As shown in Figure 2c, the hotspot counties for eclipse tweets were mostly counties on the POT after adjustment by county population. The only exception was four counties in North Dakota. Considering the noisy nature of tweet data, the hotspot county analysis results captured the eclipse event dynamics surprisingly well. Note that the hotspot analysis results did not indicate the absolute number of eclipse watchers in a county; rather they indicated the degree of concentration of eclipse watchers in a county in comparison with surrounding counties. Moreover, no county was identified as a cold spot. In other words, there was no statistically significant cluster of counties that have a lower number of eclipse tweets than their neighboring counties. In terms of absolute number of eclipse tweets, Tennessee had the most, followed by South Carolina and Missouri. For eastern states that have high population density, even if a county was not in the POT, there could still have been a large number of eclipse watchers, whereas for western states with lower population density, counties that were not on POT had significantly fewer eclipse watchers than POT counties. As a result, the most significant hotspot counties were POT counties from the western states. In eastern POT states, there were two hotspot regions. The first was bordering counties in Kentucky, Tennessee, and Illinois, including places like Nashville, TN. The second was bordering counties of Tennessee, Georgia, North Carolina and South Carolina. This reflected the movement choices of out-of-state eclipse watchers; and border counties were more likely to be the destination choices for eclipse viewing.

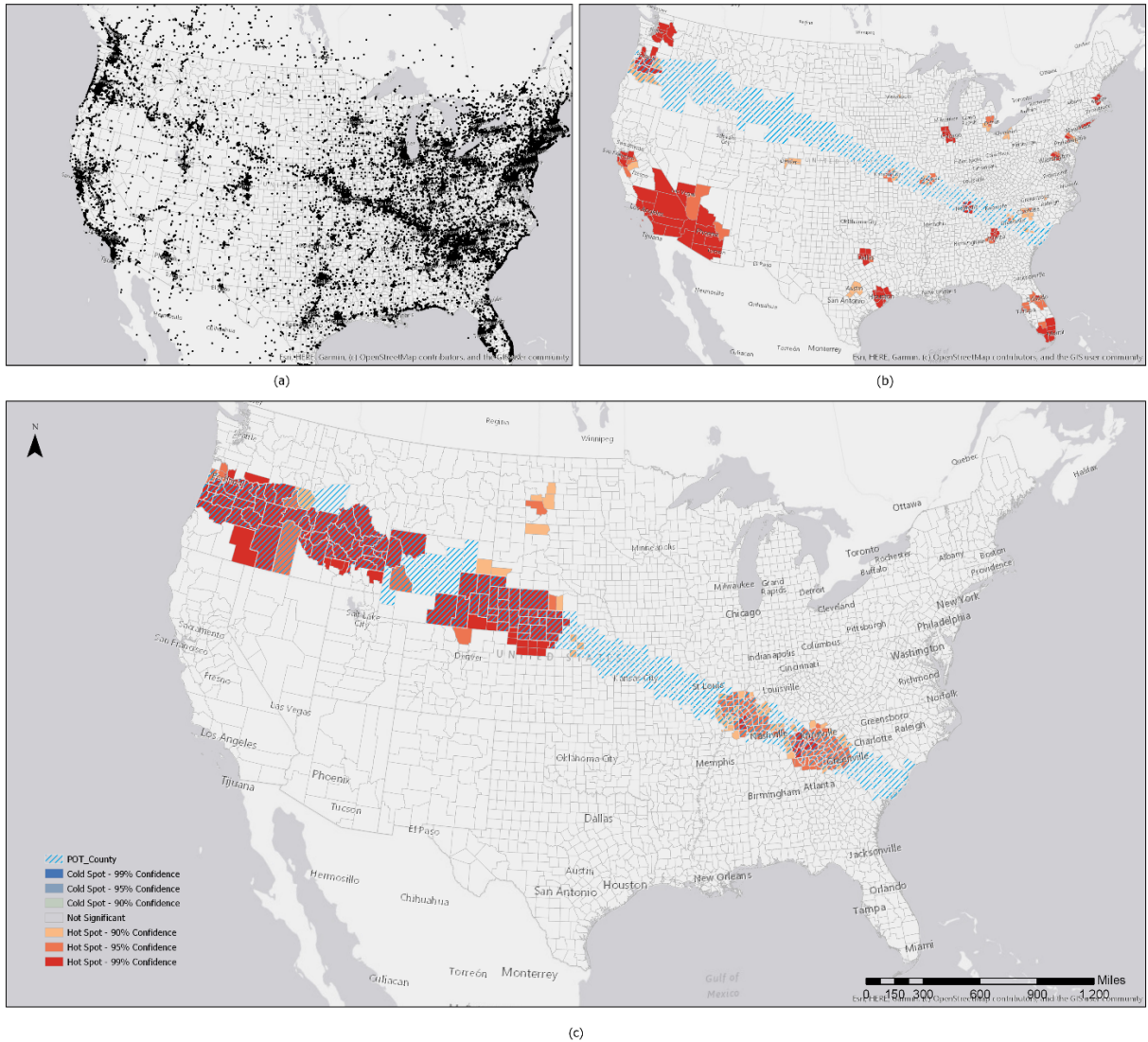

*Figure 2. Spatial distribution of eclipse tweets. (a) spatial distribution of eclipse tweets in the United States; (b) hotspot counties for eclipse tweets; (c) hotspot counties for eclipse tweets after normalizing the number of eclipse tweets by county population*

The hourly temporal distribution of eclipse tweets inside and outside the POT from Aug 20th to Aug 23rd is shown in Figure 3. The time of day information for each eclipse tweet was converted to local time zone based on the location of the eclipse tweet. The hourly distribution of eclipse tweets showed that the vast majority of eclipse tweets were posted on the eclipse day, Aug 21, proving that most of the eclipse tweets were truly in-situ sensed data. Inside the POT, the peak hour of eclipse tweets occurred at the 13th hour, which was when the totality started in the Central

Time Zone and where the Tennessee hotspot counties were. Interestingly, outside of the POT, there was a small peak at the 10th hour, when the totality first started on the west coast; and the peak hour of eclipse tweets there occurred at the 14th hour when the totality started in all the Eastern states.

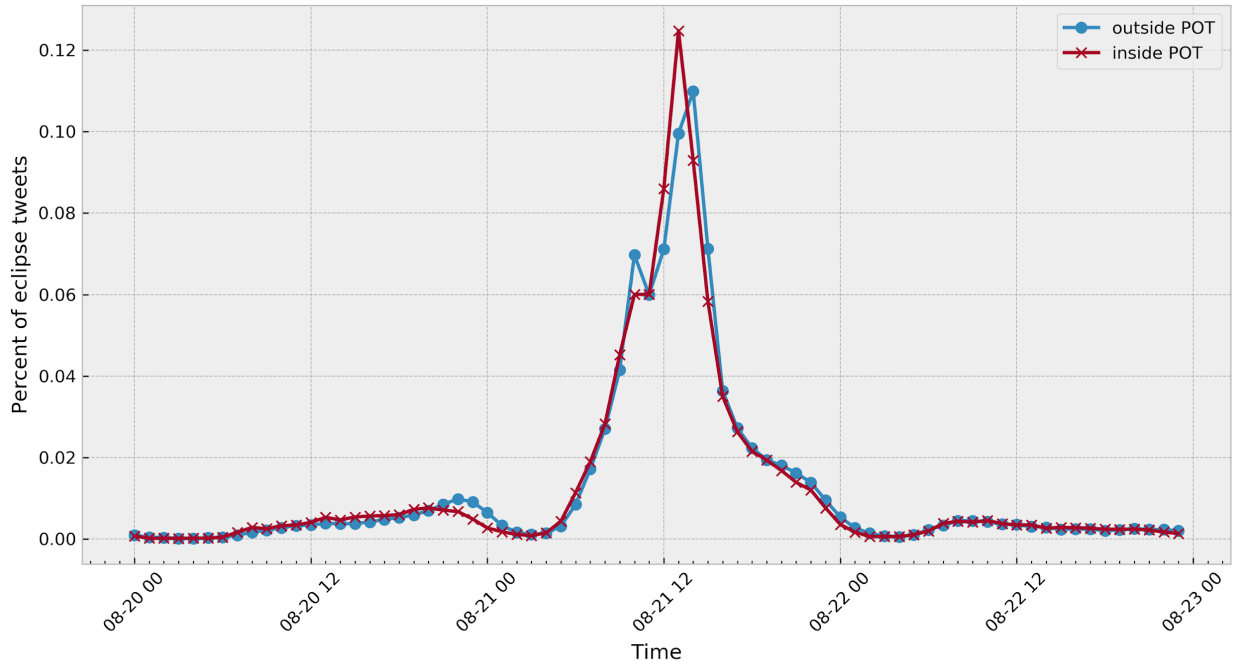

*Figure 3. Hourly fluctuation of eclipse tweets inside (red) and outside (blue) POT*

The hourly fluctuation of eclipse tweets clearly reflected the response of the population as the solar eclipse moved from west to east. In addition, the in-situ nature of the eclipse tweets indicated that the location information of eclipse tweets captured the twitter user's location while they were watching the eclipse. If we can infer eclipse watchers' home locations, then we would be able to uncover the overall movement dynamics on eclipse day based on the origin and destination information.

### 3.2 Retrieving tweet histories of eclipse watchers

The hotspot counties for eclipse tweets reveal that a large number of out-of-state visitors travelled to POT states, but could not show where these out-of-state visitors were from. From a transportation planning perspective, understanding the origins of visitors can greatly help transportation agencies to manage congestion on different roads and predict roadways that are more likely to be subject to severe traffic volume increases and plan accordingly. For *TweetMovementFlow,* we used the tweet histories of eclipse watchers to infer their home states. Compared with previous methods, using tweet histories can significantly augment the signals of users' historical spatial footprints. As Twitter users' spatial footprint histories reflect their activity spaces in real life, the region with the most frequent tweet activities was designated as a Twitter user's home region. The tweet histories of all the POT eclipse watchers were collected using Twitter user timeline API[2], which permits developers to retrieve a collection of the most recent tweets (about 3200) posted by a Twitter user. Table 1 shows some statistics of the retrieved tweet histories, which included 110,178,814 tweets from 212,696 eclipse watchers[3].

*Table 1. Statistics of tweet histories of eclipse watchers*

|  | total number of collected historical tweets | number of place-referenced tweets | number of coordinate-referenced tweets | ratio of place-referenced tweets | ratio of coordinate-referenced tweets |
|---|---|---|---|---|---|
| max | 3819 | 3245 | 3244 | 100% | 100% |
| min | 1 | 1 | 0 | 0.03% | 0 |
| median | 3055 | 345 | 1 | **17%** | 0.06% |
| mean | 2456 | 518 | 102 | **22%** | 5% |

As shown in Table 1, about 20% (both mean and median) of tweets from the tweet histories of eclipse watchers were place-referenced tweets. On average, only about 5% of the tweets in their tweet histories had coordinate information. Moreover, the median ratio of coordinate-referenced tweets (0.06%) showed that the coordinate-referenced tweets were highly skewed among different Twitter users, and some Twitter users did not have any coordinate-referenced tweets. Therefore, only using coordinate-referenced tweets to infer Twitter user's origin state was not a feasible solution as coordinate-referenced tweets were too scarce to provide enough support for reasoning. Since the ratio of *place-referenced tweets* in user tweet histories was much higher than that of *coordinate-referenced tweets*, a data fusion method was designed to integrate coordinate-referenced tweets with place-referenced tweets, further augmenting the spatial footprint signals of eclipse watchers.

---

[2] https://developer.twitter.com/en/docs/tweets/timelines/api-reference/get-statuses-user_timeline.html
[3] A small portion of Twitter users didn't authorize developers to collect their tweet histories.

**3.3 Fusing multi-scale georeference information using VB-KDE**

Because the georeferenced information of place-referenced tweets can have different spatial scales (e.g., state, city, neighborhood, poi) (Figure 4), the estimation results for different eclipse watcher's home region have different levels of uncertainty. Figure 4(a) and (b) show the spatial footprint histories of two individual Twitter users; Twitter user A posted georeferenced tweets at point, city and state scale, while Twitter user B posted only city and state-scale georeferenced tweets.

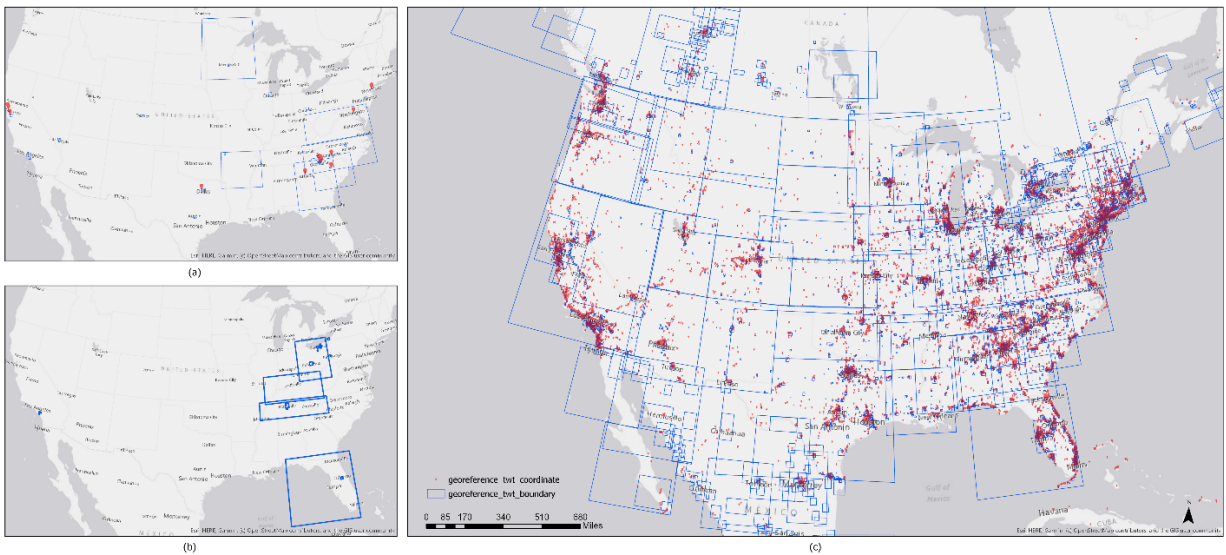

*Figure 4 (a) spatial footprint history of individual Twitter user A; (b) spatial footprint history of individual Twitter user B; (c) spatial footprint histories of randomly sampled 10 percent of eclipse watchers*

To account for the different levels of uncertainties in the georeferenced information of tweets, we designed a variable bandwidth kernel density estimation (VB-KDE) method. For traditional point kernel density estimation method, bandwidth selection is an important step. For example, ESRI's ArcGIS calculates the bandwidth for point kernel density estimation using the following formula:

$$BW_s = 0.9 * min(SD, \sqrt{\frac{1}{ln(2)} * D_m}) * n^{-0.2}$$

This formula is based on Silverman's rule-of-thumb bandwidth estimation formula but has been adapted for two dimensions (Silverman 1986). *SD* refers to standard distance, *Dm* was the median distance from all points to the mean center, and *n* is the number of points. If all georeferenced tweets were coordinate-referenced, then the ArcGIS formula can be adopted to select bandwidth. In VB-KDE, the bandwidth for georeferenced tweets varied with the scale of the georeferenced information. $BW_s$ was selected to be the standard bandwidth for coordinate-referenced tweets, and the bandwidth for place-referenced tweets was scaled according to the area of the place reference. The specific equation is:

$$BW_p = \sqrt{\frac{Area_p + \alpha}{\alpha}} * BW_s$$

where $Area_p$ was calculated from the boundary coordinates of the place reference and the unit of $Area_p$ was $m^2$. For a coordinate-referenced tweet, $Area_p$ was zero, i.e., $BW_p = BW_s$. $\alpha$ was an adjustment factor to tune the scaling effect, that was related to the spatial accuracy of point data. For coordinate-referenced tweets, the spatial accuracy was 50~100 m; $\alpha$ was set to 80 in this study.

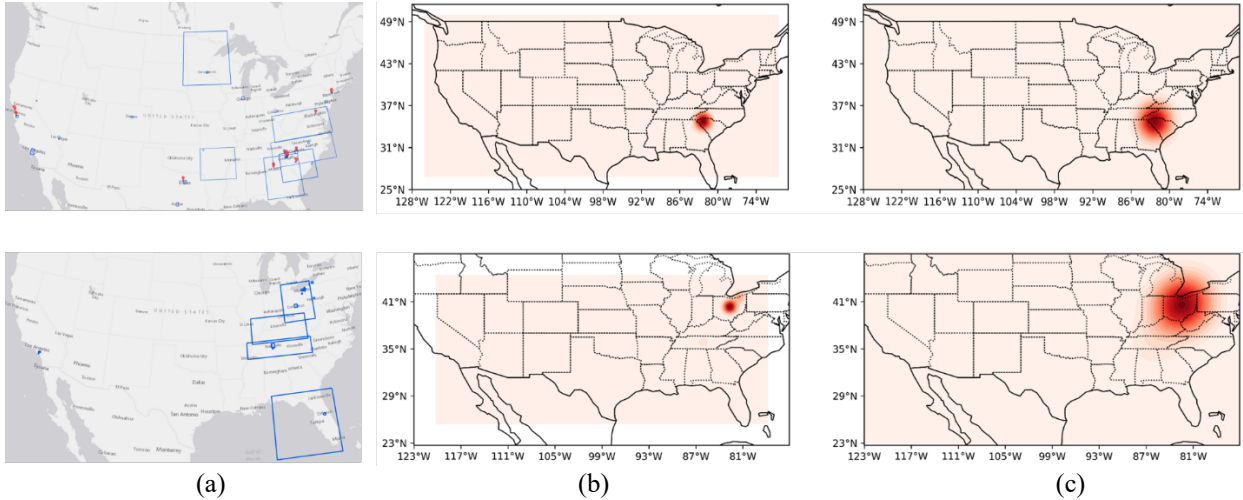

(a) (b) (c)

*Figure 5 Variable bandwidth kernel density estimation result for Twitter user A and B.*

Figure 5(c) shows the VB-KDE results for two sampled Twitter users from Figure 4(a) (b). As a comparison, Figure 5(b) also showed the KDE results using the same bandwidth, $BW_s$, for all georeferenced tweets (place-referenced tweets were represented by the center of the place boundary). As shown in Figure 5, VB-KDE results capture more of the uncertainty of coarse-scale georeferenced data.

By applying VB-KDE on each eclipse watcher's spatial footprint history, a probability activity surface was generated for each eclipse watcher. By overlaying the state shapefiles with probability activity surfaces, map algebra was used to calculate the state with the highest probabilities for each POT eclipse watchers, which was then set to be the POT eclipse watcher's origin state. To do a preliminary validation of VB-KDE method, we collected the profile location information shared by eclipse watchers and compared it with VB-KDE results. Profile location is a piece of information voluntarily shared by Twitter users as their home location, and it can be any text. The comparison shows the vast majority of profile locations matched our estimation results on a state scale. Table 2 shows the top 10 most frequent profile locations for eclipse watchers from eight POT states.

*Table 2. Comparison of profile locations and VB-KDE results for eclipse watchers*

| **Oregon** | **Idaho** | **Wyoming** | **Nebraska** | **Missouri** | **Tennessee** | **North Carolina** | **South Carolina** |
|---|---|---|---|---|---|---|---|
| "Portland, OR"; "Oregon, USA"; "Portland, Oregon"; "Eugene, OR"; "Oregon"; "Corvallis, OR"; "Bend, OR"; "Salem, OR"; "Portland"; "PDX"; | "Boise, ID"; "Idaho, USA"; "Boise, Idaho"; "Idaho"; "Meridian, ID"; "United States"; "Nampa, ID"; "Boise"; "Pocatello, ID"; "Idaho Falls, ID"; | "Wyoming, USA"; "Laramie, WY"; "Wyoming"; "Jackson, WY"; "Casper, WY"; "Cheyenne, WY"; "Casper, Wyoming"; "Jackson Hole, Wyoming"; "United States"; "Rock Springs, Wyoming"; | "Omaha, NE"; "Lincoln, NE"; "Nebraska, USA"; "Nebraska"; "Lincoln, Nebraska"; "Omaha, Nebraska"; "Omaha"; "Kearney, NE"; "Grand Island, NE"; "United States"; | "Kansas City, MO"; "St Louis, MO"; "St. Louis, MO"; "Missouri, USA"; "Kansas City"; "St. Louis"; "Columbia, MO"; "Springfield, MO"; "St. Louis, Missouri"; "Missouri"; | "Nashville, TN"; "Knoxville, TN"; "Memphis, TN"; "Tennessee, USA"; "Chattanooga, TN"; "Nashville"; "Tennessee"; "Murfreesboro, TN"; "Franklin, TN"; "Nashville, Tennessee"; | "Charlotte, NC"; "Raleigh, NC"; "North Carolina, USA"; "North Carolina"; "Durham, NC"; "Asheville, NC"; "Greensboro, NC"; "Chapel Hill, NC"; "NC"; "Wilmington, NC"; | "Charleston, SC"; "South Carolina, USA"; "Columbia, SC"; "Greenville, SC"; "South Carolina"; "Myrtle Beach, SC"; "Lexington, SC"; "United States"; "Clemson, SC"; "Columbia, South Carolina"; |

Without other available ground truth data, this comparison demonstrated the validity of VB-KDE in estimating eclipse watcher's origin state from his/her spatial footprint history.

**3.4.** Inflow of eclipse watchers from different states to each POT state

Like all sampled studies, the representativeness of a sample is critical. In the hotspot county analysis of eclipse tweets, county population has proven to be an effective correcting factor for the spatial sampling bias of tweets. Therefore, after estimating the origin states of POT eclipse watchers, the number of eclipse watchers from each state were calibrated by state population, which was used as a weighting factor, the higher the state population, the lower the weight. Then, the weighted number of each state was divided by the weighted sum to calculate the percentage of eclipse watchers from different states. In the following discussion, unless otherwise specified, the number of eclipse watchers refers to the calibrated number. In total, 28,596 POT eclipse watchers were collected as our study sample. For each POT state, the emphasis of our analysis was on the ranked order of percent of eclipse watchers from different states rather than the absolute number of eclipse watchers. In other words, the goal was to understand where most eclipse watchers were from for each POT state. Figure 6 shows the percent of eclipse watchers from different states for each of the eight states that the POT passed through.

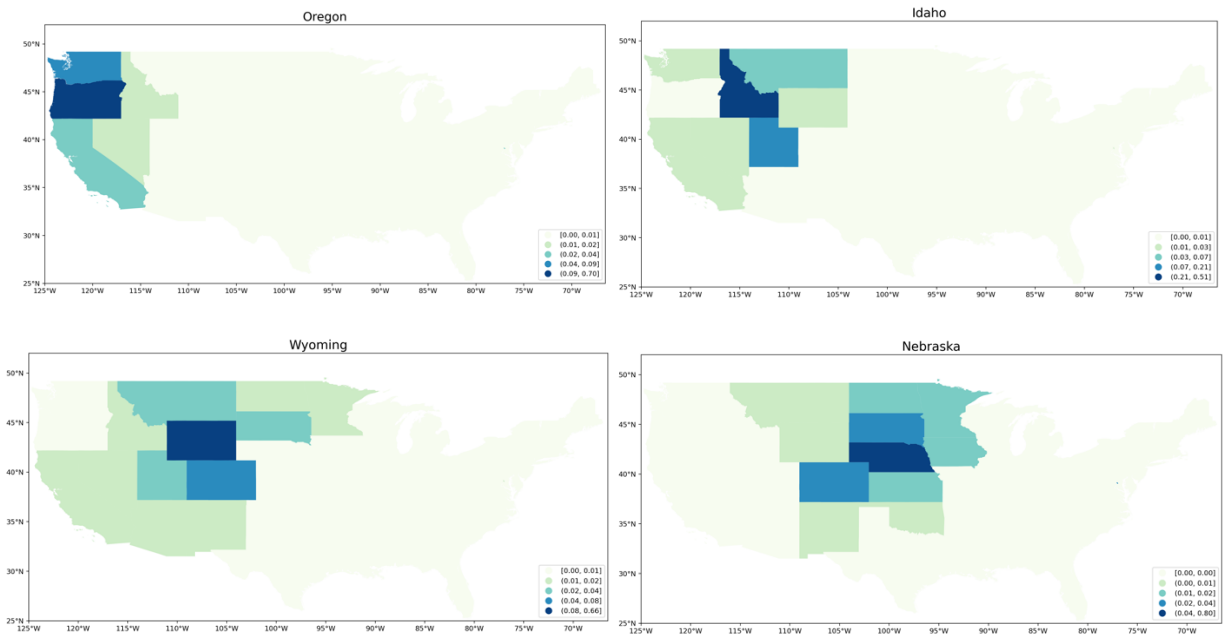

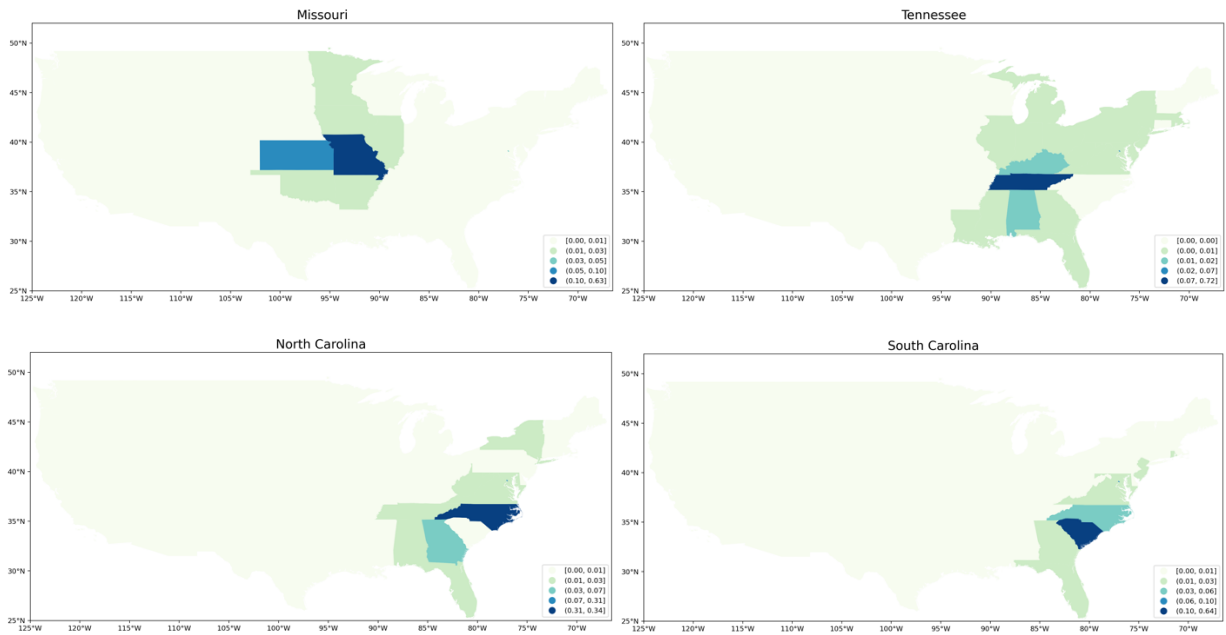

*Figure 6 Percent of POT eclipse watchers from different states in eight POT states*

As shown in Figure 6, the majority of eclipse watchers (60%~70%) in each POT state were from within the state, and most out-of-state eclipse watchers were from neighboring states, especially neighboring states that were not in POT. Oregon attracted the most out-of-state eclipse watchers from Washington and California. Most out-of-state eclipse watchers travelling to Idaho were from Utah and Montana, while very few eclipse watchers travelled from Oregon to Idaho. Most out-of-state eclipse watchers in Wyoming were from Colorado. For Nebraska, South Dakota and Colorado contributed the most out-of-state eclipse watchers. Missouri attracted most of its out-of-state eclipse watchers from Kansas, while Tennessee attracted most from Kentucky and Alabama. North Carolina attracted most out-of-state eclipse watchers from Georgia but not South Carolina; South Carolina attracted most of its out-of-state eclipse watchers from North Carolina, maybe because South Carolina was the last state from which the solar eclipse was visible. Except for South Carolina, POT states attracted most of their eclipse watchers from neighboring states that were not on the POT. To validate that this traveler pattern was indeed due to the impact of the solar eclipse event, we will compare it with traveler patterns during a more regular time period in the next section.

## 4. Results validation

To validate the effectiveness of the *TweetMovementFlow* analytic framework, we first compared the estimation results with the official traffic count reports released by State Transportation agencies for two states that released their eclipse day traffic count data, namely, Idaho and Wyoming. Our estimation results showed that the *largest number of out-of-state visitors to Idaho on the eclipse day was from Utah*. This matches the Idaho State Transportation Department's traffic count report for eclipse day, which estimated more than 160,000 visitors came from out of state, and the busiest road on the eclipse day was the I-15 corridor between Utah and Idaho Falls[4] (Figure 7). For Wyoming, our method estimated that *most out-of-state visitors were from Colorado, followed by Uta*h. This also matches with the traffic count report released by Wyoming DOT. According to the traffic count report released by Wyoming DOT, Laramie County located on the southeastern border between Wyoming, Colorado had the largest increase in traffic for eclipse day and traffic on I-25 in southern Wyoming increased the most on eclipse day[5]. For traffic on I-80, WYDOT reported that I-80 east of Evanston (on the border of Utah and Wyoming) had the most significant traffic increase (Figure 7).

---

[4] https://itd.idaho.gov/news/solar-eclipse-traffic-counts/
[5] http://www.dot.state.wy.us/news/wyoming-traffic-increased-monday-by-more-than-536000-during-aug-21-sol

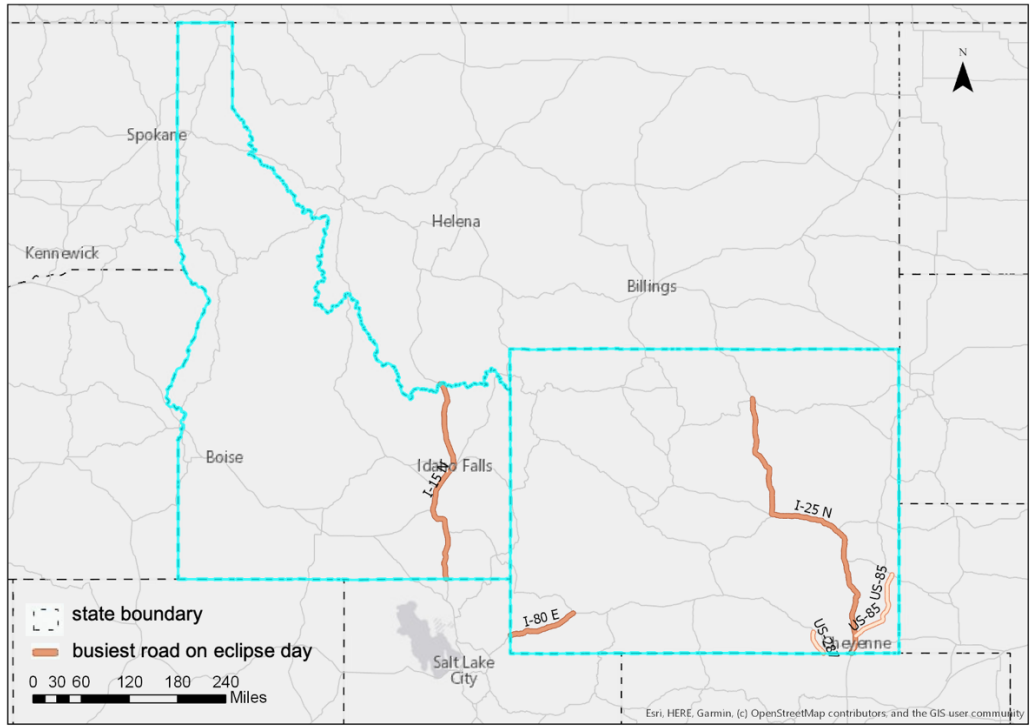

*Figure 7 Roads in Idaho and Wyoming with the highest traffic volume increase on eclipse day*

To further validate that the derived movement patterns during the solar eclipse event were indeed a reflection of population response to the event, not regular spatial interactions among neighboring states, we selected Oregon as an example and compared the derived patterns with patterns in a regular time period, i.e., two weeks prior to the eclipse week (08/04/2017~08/17/2017). For this period, we identified 17,044 Twitter users who have visited Oregon, that is, Twitter users who have tweeted in Oregon. These Twitter users were considered as a sample of *regular visitors* to Oregon; over 34 million historical tweets were collected to infer regular visitors' home states using the VB-KDE method.

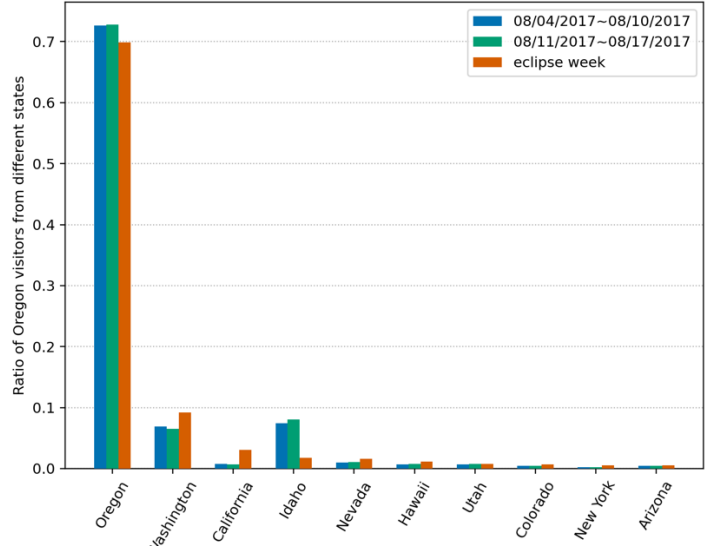

*Figure 8 Top 10 home states for Oregon visitors during regular time and the eclipse week*

As shown in Figure 8, two weeks prior to the solar eclipse event, the four neighboring states of Oregon, Idaho, Washington, Nevada and California contributed the most visitors to Oregon. In contrast, the ratio of visitors in Oregon that were from Idaho dropped significantly during eclipse week, while the percent of the other three neighboring states increased the most during the eclipse week. The change demonstrated that georeferenced tweets indeed captured human movement dynamics in response to a large-scale event and the proposed analytic framework was able to extract the movement pattern.

## 5. Sampling bias and fusion of multi-scale geosocial data

Social media data, particularly geosocial data is a noisy and biased sample, but it provides a unique perspective for researchers to explore different aspects of human dynamics. The sampling bias of geosocial can be separated into spatial bias and temporal bias. Spatial bias is primarily caused by the underlying population distribution, and research has demonstrated the bias towards urban population center (Johnson et al., 2016). In this paper, we used population data as an adjustment factor to correct the spatial sampling bias of tweet data from different states. Although the correction effect is not perfect, population size can effectively mitigate the spatial sampling bias largely.

Moreover, spatial and temporal closeness to the event is more likely to stimulate people to provide in-situ response, as demonstrated by spatiotemporal analysis of eclipse tweets. This self-selection process becomes almost a positive discrimination factor (Ciulla et al. 2012) for using geosocial data to study collective human movement dynamics. In comparison with traditional data sources, the real-time, in-situ, shallow but wide nature of geosocial data is actually well suited for studying human dynamics in response to large-scale natural and social events. We can easily imagine applying the proposed big geosocial data analytic framework to other scenarios such as hurricane evacuations, disease outbreaks, where large-scale real-time data sources that have big enough coverage are not yet available.

As we enter the era of Internet of Things, different location-aware devices (e.g., mobile phone, GPS tracker, RFID sensor) have become ubiquitous, and spatiotemporal footprints of moving entities will be one of the most fundamental types of data that researchers have to handle. The fusion of spatial data from different sources and with different spatial scales will be a critical step in the data mining process. The different scales of georeferenced information associated with tweets is an example of data with varying scales; the proposed data fusion approach can be easily applied to other geographic data sources.

## 6. Conclusions and future work

In this research, we developed an analytic framework, *TweetMovementFlow*, to extract human movement dynamics in response to a large-scale event from big geosocial data. Using the 2017 Great American Solar Eclipse as an example event, we demonstrated that big geosocial data could capture the real-time response of population to large-scale events, and the proposed analytic framework can effectively extract human movement dynamics from big geosocial data. In view of the unique characteristics of geosocial data, we developed several approaches to mitigate the data scarcity and sampling bias issues of geosocial data, including the two-stage data collection process and data aggregation and calibration. We also developed a VB-KDE approach to fuse multi-scale georeference data. Georeferenced tweets with different spatial scales have different levels of uncertainty. VB-KDE approach can handle the changing uncertainties through varying bandwidth. Researchers who study different aspects of human dynamics can use these approaches to augment signals of human dynamics in socially sensed data.

The applicability of the proposed framework depends on the salience of an event, which is determined by the spatiotemporal extent and semantic uniqueness of the event. The more unusual an event is, people as social sensors are more likely to respond and tweet about it. In addition, the population and geographic region that are more directly impacted are more likely to respond and generate social signals about the event. These two unique characteristics of geosocial data make it a feasible and valuable data source for studying population response to large-scale remarkable events. Other types of large-scale events like earthquakes, hurricanes, and epidemic outbreaks have the saliency and are also likely to stimulate a great deal of social responses from the population. Different kinds of events will generate different changes in human movement dynamics. For example, natural hazard like earthquake will not give the population a lot time to plan ahead; hurricane will cause population displacement over a prolonged period of time; epidemic outbreaks will reduce people's mobility. The complicated changes of human movement dynamics, over various spatial and temporal ranges, will pose significant challenges for traditional data and analytic approaches, which further proves the value of the proposed framework. The proposed framework could be applied to analyze these types of events by adapting the first step of the data collection process, i.e., adjusting the spatial and temporal ranges of tweet collection, to collect tweets about the event of interest, which should be easy to do given the semantic saliency of the event. Other modules of the framework can be applied with minimal adaption.

For future work, several interesting topics need further study. To reduce uncertainty, we only aggregated eclipse tweets by two spatial scales, county and state, in our analysis. However, the majority of eclipse tweets have finer spatial scales, and we can potentially obtain human movement dynamics with finer spatial scale. Finding a balance between spatial scale and uncertainty in geosocial data analysis is an important topic worth further study. Moreover, solar eclipses are a relatively unique event. In future work, we plan to apply this analytic framework to different events for comparison.

## Acknowledgement

This research has been supported in part by a grant from the Maryland Transportation Institute.

Balcan, D., Colizza, V., Gonçalves, B., Hu, H., Ramasco, J. J., & Vespignani, A. (2009). Multiscale mobility networks and the spatial spreading of infectious diseases. *Proceedings of the National Academy of Sciences of the United States of America*, *106*(51), 21484–21489.

Brambilla, M., Ceri, S., Daniel, F., & Donetti, G. (2017). Spatial Analysis of Social Media Response to Live Events. In *Proceedings of the 26th International Conference on World Wide Web Companion - WWW '17 Companion*. https://doi.org/10.1145/3041021.3051698

Cheng, Z., Caverlee, J., & Lee, K. (2010). You Are Where You Tweet: A Content-based Approach to Geo-locating Twitter Users. *Proceedings of the 19th ACM International Conference on Information and Knowledge Management*, 759–768.

Ciulla, F., Mocanu, D., Baronchelli, A., Gonçalves, B., Perra, N., & Vespignani, A. (2012). Beating the news using social media: the case study of American Idol. *EPJ Data Science*, *1*(1), 8.

Davis, C. A., Jr., Pappa, G. L., de Oliveira, D. R. R., & de L. Arcanjo, F. (2011). Inferring the Location of Twitter Messages Based on User Relationships. *Transactions in GIS*, *15*(6), 735–751.

Eubank, S., Guclu, H., Kumar, V. S. A., Marathe, M. V., Srinivasan, A., Toroczkai, Z., & Wang, N. (2004). Modelling disease outbreaks in realistic urban social networks. *Nature*, *429*(6988), 180–184.

Fan, J., Fu, C., Stewart, K., & Zhang, L. (2019). Using big GPS trajectory data analytics for vehicle miles traveled estimation. *Transportation Research Part C: Emerging Technologies*, *103*, 298–307.

González, M. C., Hidalgo, C. A., & Barabási, A.-L. (2008). Understanding individual human mobility patterns. *Nature*, *453*(7196), 779–782.

Guan, W., Gao, H., Yang, M., Li, Y., Ma, H., Qian, W., Cao, Z., & Yang, X. (2014). Analyzing user behavior of the micro-blogging website Sina Weibo during hot social events. *Physica A: Statistical Mechanics and Its Applications*, *395*, 340–351.

Hecht, B., & Stephens, M. (2014). A tale of cities: Urban biases in volunteered geographic information. *Eighth International AAAI Conference on Weblogs and Social Media*. https://www.aaai.org/ocs/index.php/ICWSM/ICWSM14/paper/viewPaper/8114

Huang, Q., & Wong, D. W. S. (2015). Modeling and Visualizing Regular Human Mobility Patterns with Uncertainty: An Example Using Twitter Data. *Annals of the Association of American Geographers. Association of American Geographers*, *105*(6), 1179–1197.

Johnson, I. L., Sengupta, S., Schöning, J., & Hecht, B. (2016). The Geography and Importance of Localness in Geotagged Social Media. *Proceedings of the 2016 CHI Conference on Human Factors in Computing Systems*, 515–526.

Jurdak, R., Zhao, K., Liu, J., AbouJaoude, M., Cameron, M., & Newth, D. (2015). Understanding Human Mobility from Twitter. *PloS One*, *10*(7), e0131469.

Jurgens, D., Finethy, T., McCorriston, J., Xu, Y. T., & Ruths, D. (2015). Geolocation prediction in twitter using social networks: A critical analysis and review of current practice. *Ninth International AAAI Conference on Web and Social Media*. https://www.aaai.org/ocs/index.php/ICWSM/ICWSM15/paper/viewPaper/10584

Kong, L., Liu, Z., & Huang, Y. (2014). SPOT: Locating Social Media Users Based on Social Network Context. *Proceedings of the VLDB Endowment International Conference on Very Large Data Bases*, *7*(13), 1681–1684.

Kulshrestha, J., Kooti, F., Nikravesh, A., & Gummadi, K. P. (2012). Geographic dissection of the twitter network. *Sixth International AAAI Conference on Weblogs and Social Media*. https://www.aaai.org/ocs/index.php/ICWSM/ICWSM12/paper/viewPaper/4685

Liu, J., Zhao, K., Khan, S., Cameron, M., & Jurdak, R. (2015). Multi-scale population and mobility estimation with geo-tagged Tweets. *2015 31st IEEE International Conference on Data Engineering Workshops*, 83–86.

Lloyd, A., & Cheshire, J. (2017). Deriving retail centre locations and catchments from geo-tagged Twitter data. *Computers, Environment and Urban Systems*, *61*, 108–118.

Lu, X., Bengtsson, L., & Holme, P. (2012). Predictability of population displacement after the 2010 Haiti earthquake. *Proceedings of the National Academy of Sciences of the United States of America*,
21